# A Fast fixed-point Quantum Search Algorithm by using Disentanglement and Measurement

Ashish Mani* and C. Patvardhan

*Abstract*—Generic quantum search algorithm searches for target entity in an unsorted database by repeatedly applying canonical Grover's quantum rotation transform to reach near the vicinity of the target entity. Thus, upon measurement, there is a high probability of finding the target entity. However, the number of times quantum rotation transform is to be applied for reaching near the vicinity of the target is a function of the number of target entities present in an unsorted database, which is generally unknown. A wrong estimate of the number of target entities can lead to overshooting or undershooting the targets, thus reducing the success probability. Some proposals have been made to overcome this limitation. These proposals either employ quantum counting to estimate the number of solutions or fixed-point schemes. This paper proposes a new scheme for stopping the application of quantum rotation transformation on reaching near the targets by disentanglement, measurement and subsequent processing to estimate the distance of the state vector from the target states. It ensures a success probability, which is greater than half for all practically significant ratios of the number of target entities to the total number of entities in a database. The search problem is trivial for remaining possible ratios. The proposed scheme is simpler than quantum counting and more efficient than the known fixed-point schemes. It has same order of computational complexity as canonical Grover`s search algorithm but is slow by a factor of two and requires two additional ancilla qubits.

## I. INTRODUCTION

Quantum Computation has emerged as an exciting field in recent past [1] with development of Shor`s polynomial time Factorizing algorithm [2] and Grover`s Quantum Search algorithm [3]. Grover`s Quantum Search algorithm finds wide application as a subroutine in solving many types of problems [4]. Grover`s Algorithm is superior to classical search algorithms as it provides for quadratic speed up in search process. It is designed for searching a unique item from an unsorted database containing N items [3]. It is assumed that an oracle exists that informs upon querying whether a selected item satisfies the criterion in one-step. Classically, this problem requires an average of N/2 such oracle calls. Grover's algorithm can find the unique item in $O(\sqrt{N})$ steps, which is quadratic speedup as compared to any known classical algorithm. Grover's algorithm has been further studied in detail in [5] and tighter bounds have been provided for the quantum search. It is observed that the quantum search takes place by rotating the state vector in a two dimensional Hilbert subspace associated with the database. If the application of rotation transform is not stopped near the target state, the state vector would either undershoot or overshoot the

Contact: ashish.mani@ieee.org

target state considerably. Further, it has been established that the number of times rotation transform is required to be applied on the state vector is a function of not only the size of the database but also the number of target states. The number of target states is mostly unknown for real world cases, thus, it limits the application of the canonical quantum search algorithm. A number of attempts have been made to solve this problem ranging from employing quantum counting to estimate the number of target states [5] to fixed point quantum searches [6] and critically damped quantum searching [7].

In this paper, a novel scheme has been suggested to solve this problem by employing disentanglement, measurement and subsequent processing to estimate the distance of the state vector from the target states. This paper integrates the idea of indirect measurement from instrumentation to estimate the distance from the target state and feedback control to stop near the target state.

The paper is further organized as follows: Section 2 presents existing quantum search algorithms. Section 3 describes and analyzes the proposed algorithm. Conclusions are drawn in Section 4.

## II. EXISTING QUANTUM SEARCH ALGORITHMS

The canonical quantum search algorithm proposed by Grover [3] solves a general search problem in which there are '$N$' elements that can be represented by '$n$' basis states in Hilbert Space i.e. $N \leq 2^n$, where '$N$' and '$n$' are both positive integers. Let $H_S = \{0, 1\}^n$ and let $O_f : H_S \rightarrow \{0, 1\}$, where $O_f$ represents the oracle, which returns the answer when sampled but no other information is known about $O_f$. Using this framework along with quantum operators, the target state '$t_s$', which is a basis state in $H_S$ such that $O_f(t_s) = 1$, is to be found. It consists of the following steps:

(i) Initializing a set of qubits $|S\rangle$, which represent the solutions and an output qubit.

(ii) Apply quantum rotation gate, G, 'r' times on the qubits to amplify the probability of finding target element.

(iii) Measure solution vector $|S\rangle$.

(iv) If $|S\rangle$ contain the correct information, then stop.

(v) Else, go to step (i).

Grover`s algorithm is a probabilistic algorithm with computational complexity of O(√N) and its performance has been studied in [5]. It has been proved to be optimal up to a constant factor. The probability of finding the target element is a function of 'r' given by the following equation [4]:

$$g_r(P) = \sin^2((2r+1)\sin^{-1}(\sqrt{P})) \qquad (1)$$

where P = m/N, 'm' is the number of target states / elements and N is the total number of elements. It is evident that if 'm' is known in advance then the target elements can be easily located. However, 'm' is mostly not known, in advance, in real life problems. Further, if 'r' is selected smaller or even larger than the ideal value, the probability of finding the target elements can reduce considerably. Thus, the probability of finding the solution is reduced not only when less number of quantum rotations are performed but also when more number of quantum rotations are performed than the optimal.

A number of ways have been devised to overcome this limitation of canonical quantum search algorithm, which includes amplitude amplification and quantum counting [5], intelligent heuristic guesses [8-9], Fixed-point algorithms [10], measurement based fixed-point algorithm [6], Critically damped quantum search [7].

Amplitude amplification and quantum counting techniques increases the requirement of number of different types of quantum operators, which may be practically more difficult to implement, at least initially, and further they would require additional

Contact: ashish.mani@ieee.org

queries [5].

Fixed-point algorithms proposed in [10] makes the quantum search algorithm behave in a similar way as classical search algorithm i.e. Fixed point algorithm does not overshoot the target state and moves monotonically towards it. However, the computational complexity of the proposed algorithm is of O(N) though it is faster by a factor of 2 in comparison to the classical algorithm and is also proven to be asymptotically optimal [6].

Fixed-point quantum algorithms based on measurement have also been suggested in literature [6] which improves on the algorithm proposed in [10]. However, it still has the same computational complexity i.e. O(N) rather than the optimal O($\sqrt{N/m}$) of canonical quantum search algorithm. The fixed-point algorithms have been developed for the problems in which expected number of queries is small.

This paper proposes an improved measurement based fixed-point algorithm that provides the same computational complexity as that of canonical quantum search i.e. O($\sqrt{N/m}$) irrespective of the number of target entities by disentangling, measuring the ancilla qubit and subsequently counting the number of times $|0\rangle$ and $|1\rangle$ have been obtained to compute their ratio. This ratio is used for determining the continuing and stopping criterion of application quantum rotation transform on $|S\rangle$ qubits.

Recently, a critically damped quantum search algorithm described in [7] has been proposed, which solves the problem by introducing damping in Grover's step in O($\sqrt{N/m}$), whereas the proposed algorithm solves the problem externally, mostly with systems perspective i.e. by integrating ideas and establishing a mathematical relationship without bringing any change in the existing Grover's Step. However, they both use similar feedback during search process but the processing involved in feedback is different. The algorithm described in [7] has two versions: first one in which there is a static value of $\cos\varphi = (1-\sin\theta)/(1+\sin\theta)$ is used and works well for m<<N. However, for large m, 1.5 times of Grover's canonical search algorithm for searching unique element in database, is a touch slow so the second version was proposed with dynamic $\cos\varphi_n = (1-\sin\pi/2n)/(1+\sin\pi/2n)$, which relies heavily on flipping of external spin. However, there is a tradeoff between the quantum character of the search, which depends on the value of φ and the ability of the external spin to act as an accurate indicator that the target has been found. That is, when φ increases, the nature of search also shifts from Quantum to classical, and the external spin starts giving better indication of reaching the target. Therefore, the proposed algorithm is simpler, faster and more stable than critically damped search algorithm as there is no tradeoff involved in optimization of parameters like φ.

III. PROPOSED ALGORITHM

The proposed algorithm is illustrated in Fig. 1. It employs $|S\rangle$ qubits whose basis states represent entities in a database. It also includes two ancilla output qubits, $|OQ_1\rangle$ and $|OQ_2\rangle$ and a blank qubit. The Grover's canonical quantum rotation gate 'G' and an additional oracle query transformation 'Of1' along with Local Cloning Transformation (LTC) operator for disentanglement and measurement operators for $|S\rangle$ and $|OQ_2\rangle$ have been used. Further, two classical counters, $C_1$ and $C_0$, for counting the number of times $|OQ_2\rangle$ collapses to $|1\rangle$ and $|0\rangle$ respectively have been used. The ratio $C_1/C_0$ has been employed for estimating the distance of $|S\rangle$ from the target states. The proposed algorithm would be a success if it can be shown that for a specific threshold, Set_Val, of ratio $C_1'/C_0'$, the success probability is at least greater than half, when the ratios of the number of target entities to the total number of entities are less than half in a database. The search problem is trivial for remaining possible ratios of the number of target entities to the total number of entities.

The proposed algorithm is as given below:

(i) Initialize counters for measuring the state of the output qubit-2, $|OQ_2\rangle$, post measurement to zero, i.e. $C_0$ for $|0\rangle$ and $C_1$ for $|1\rangle$.

Contact: ashish.mani@ieee.org

(ii) Initializing a set of qubits |S⟩ in |0⟩ and the ancilla qubit, |OQ$_1$⟩ in state |1⟩.

(iii) Apply Walsh-Hadamard gate on |S⟩ and |OQ$_1$⟩.

(iv) Set |OQ$_2$⟩ in state |0⟩.

(v) Apply Of$_1$ on |S⟩ and |OQ$_2$⟩.

(vi) Locally clone |OQ$_2$⟩ to disentangle it from |S⟩.

(vii) Measure the |OQ$_2$⟩ and increment C$_0$ or C$_1$ depending on the outcome of the measurement and Compute $C_1^{'}/C_0^{'}$.

(viii) Apply G on |S⟩ and |OQ$_1$⟩ to amplify the probability of finding the target elements.

(ix) If ratio $C_1^{'}/C_0^{'}$ >= Set_Val, then measure the solution qubits, else go to (iv).

(x) If the qubits contain the correct information, then stop.

Else, go to step (i).

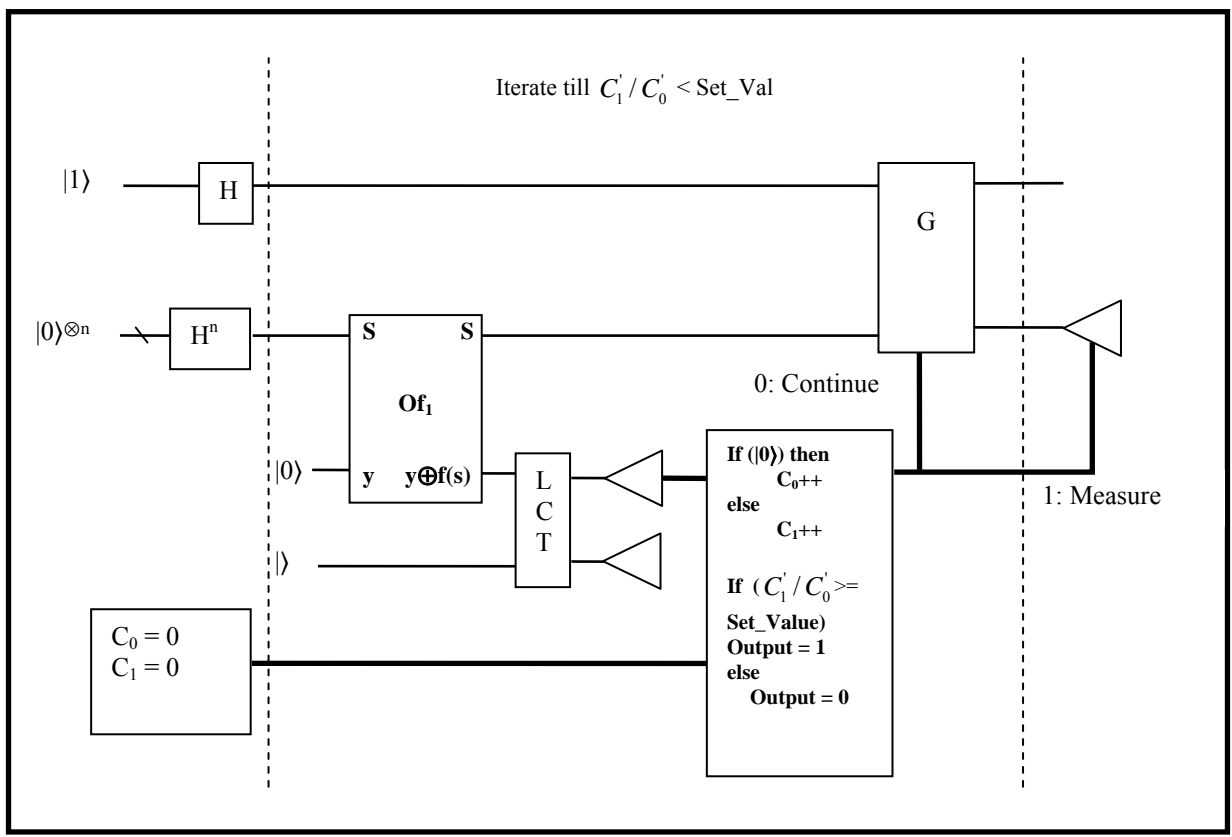

Fig. 1. Quantum circuit model of the proposed Algorithm

The main feature of this algorithm is thus, deriving a relationship between expected value of $C_1/C_0$ (i.e. $\langle C_1 \rangle / \langle C_0 \rangle$) and the success probability for all values of P less than ½. In order to derive it, let us first discuss the need for disentanglement and role of local cloning. The application of Oracle Of$_1$, entangles |S⟩ and |OQ$_2$⟩. |S⟩ can be represented in two dimensional subspace by target states |t⟩ and non-target states |t$^\perp$⟩ (orthogonal to |t⟩) by the following equation:

|S⟩ = cosθ |t$^\perp$⟩ + sinθ |t⟩,             (2)

where θ = sin$^{-1}$ (√P) and P = m/N.

Contact: ashish.mani@ieee.org

$Of_1 : |S\rangle|OQ_2\rangle \rightarrow \cos\theta\ |t^\perp\rangle|0\rangle + \sin\theta\ |t\rangle|1\rangle$ (3)

Thereby, if, $|OQ_2\rangle$ is measured, it would leave $|S\rangle$ either in the superposition of all target states or non-target states depending on the outcome of the measurement. However, if the measurement is made and discarded then it leaves $|S\rangle$ unaffected i.e. due to partial tracing, the reduced density operator of $|S\rangle$ remains same before and after the measurement. Similarly, if a measurement is made on $|S\rangle$ and discarded after applying $Of_1$, then $|OQ_2\rangle = |f(S)\rangle$, i.e. due to partial tracing, its reduced density operator pre and post measurement remains same. Since this proposal requires that both the set of qubits should be available post measurement, therefore, disentanglement is required.

Disentanglement can be defined as the process of transforming a state of two or more subsystems into an un-entangled state, which implies that the reduced density matrices of each of the subsystem remains unaffected [11]. However, it has been shown that universal disentanglement is impossible [12]. But it is still possible to disentangle a bipartite system such that the post disentanglement reduced density operators are close to the entangled reduced density operators. There are two methods available in literature for disentanglement of pure bipartite systems viz., "Disentanglement by local cloning of one qubit" and "Disentanglement by local cloning of both the qubits" [11]. In the first method, one of the subsystem is cloned to disentangle the system. The other subsystem remains unaffected, but the cloned qubit is copied rather poorly as the maximum fidelity with which it has to be copied for disentanglement is 2/3. Therefore, after disentanglement, the reduced density matrices of the subsystem are given by:

$$\rho^S_{ad} = \rho^S_{bd} \quad (4)$$

$$\rho^{OQ_2}_{ad} = \eta * \rho^{OQ_2}_{bd} + I*(1-\eta)/2 \quad (5)$$

where $\rho^S_{ad}$ is the reduced density matrix of $|S\rangle$ after disentanglement, $\rho^S_{bd}$ is the reduced density matrix of $|S\rangle$ before disentanglement, $\rho^{OQ_2}_{ad}$ is the reduced density matrix of $|OQ_2\rangle$ after disentanglement $\rho^{OQ_2}_{bd}$ is the reduced density matrix of $|OQ_2\rangle$ before disentanglement, $I$ is 2X2 identity matrix, $\eta$ is the reduction factor of the universal isotropic cloner and $\eta_{max} = 1/3$ [11].

The disentanglement by local cloning of both qubits changes the reduced density matrices of both the subsystems, so it is not as suitable for the proposed algorithm as the disentanglement by local cloning of a single qubit, which leaves one of the subsystems unaffected. The sate of qubit $|OQ_2\rangle$ post cloning is analyzed as follows.

$$\rho^{OQ_2}_{bd} = |OQ_2\rangle\langle OQ_2| \quad (6)$$

Thus, $\rho^{OQ_2}_{bd}$ can be obtained by measuring and discarding $|S\rangle$ from equation (2.3.2). Therefore,

$$\rho^{OQ_2}_{bd} = (\cos\theta|0\rangle + \sin\theta|1\rangle)(\cos\theta\langle 0| + \sin\theta\langle 1|) \quad (7)$$

$$= \begin{Bmatrix} \cos^2\theta & \cos\theta\sin\theta \\ \sin\theta\cos\theta & \sin^2\theta \end{Bmatrix}$$

Contact: ashish.mani@ieee.org

$$\rho_{ad}^{OQ_2} = \eta * \begin{Bmatrix} \cos^2\theta & \cos\theta\sin\theta \\ \sin\theta\cos\theta & \sin^2\theta \end{Bmatrix} + \left(\frac{1-\eta}{2}\right) * \begin{Bmatrix} 1 & 0 \\ 0 & 1 \end{Bmatrix} \tag{8}$$

$$\rho_{ad}^{OQ_2} = \begin{Bmatrix} \eta\cos^2\theta + \left(\frac{1-\eta}{2}\right) & \eta\cos\theta\sin\theta \\ \eta\sin\theta\cos\theta & \eta\sin^2\theta + \left(\frac{1-\eta}{2}\right) \end{Bmatrix} \tag{9}$$

Therefore, the probability of measuring $|OQ_2\rangle$ in computational basis state $|0\rangle$ =

$$Tr\left(|0\rangle\langle 0|\rho_{ad}^{OQ_2}\right) = \eta\cos^2\theta + \left(\frac{1-\eta}{2}\right) \tag{10}$$

And the probability of measuring $|OQ_2\rangle$ in computational basis state $|1\rangle$ =

$$Tr\left(|1\rangle\langle 1|\rho_{ad}^{OQ_2}\right) = \eta\sin^2\theta + \left(\frac{1-\eta}{2}\right) \tag{11}$$

Therefore, $\langle C_0 \rangle = \sum \left(\eta\cos^2\theta + \frac{1-\eta}{2}\right) \tag{12}$

And $\langle C_1 \rangle = \sum \left(\eta\sin^2\theta + \frac{1-\eta}{2}\right) \tag{13}$

The probability of finding $|S\rangle$ in non-target state = $\cos^2\theta$ (14)
And probability of finding $|S\rangle$ in target state = $\sin^2\theta$ (15)

Lets define $\langle C_0' \rangle$ and $\langle C_1' \rangle$ such that

$$\langle C_0' \rangle = \sum \cos^2\theta \tag{16}$$

$$\langle C_1' \rangle = \sum \sin^2\theta \tag{17}$$

$\langle C_0' \rangle$ can be written in terms of $\langle C_0 \rangle$ by using equations (12) and (16). Therefore,

$$\langle C_0' \rangle = \frac{1}{\eta}\left(\langle C_0 \rangle - \sum\left(\frac{1-\eta}{2}\right)\right) \tag{18}$$

$\langle C_1' \rangle$ can be written in terms of $\langle C_1 \rangle$ by using equations (13) and (17). Therefore,

$$\langle C_1' \rangle = \frac{1}{\eta}\left(\langle C_1 \rangle - \sum\left(\frac{1-\eta}{2}\right)\right) \tag{19}$$

Using equations (18) and (19), the following equation is obtained:

$$\frac{\langle C_1' \rangle}{\langle C_0' \rangle} = \frac{\langle C_1 \rangle - \sum\left(\frac{1-\eta}{2}\right)}{\langle C_0 \rangle - \sum\left(\frac{1-\eta}{2}\right)} = \frac{\sum\sin^2\theta}{\sum\cos^2\theta} \tag{20}$$

The relationship between expected value of $C_1'/C_0'$ (i.e. $\langle C_1' \rangle/\langle C_0' \rangle$) and the success probability for all values of P less

Contact: ashish.mani@ieee.org

than ½ is derived in the following way:

$$\frac{\langle C_1' \rangle}{\langle C_0' \rangle} = \frac{\sum_{r=0}^{R_n} g_r(P)}{\sum_{r=0}^{R_n} (1 - g_r(P))} \quad (21)$$

where $R_n$ is the number of rotations at the chosen value of $g_r(P)$, which should vary from 0.5 to 1.

Or $\frac{\langle C_1' \rangle}{\langle C_0' \rangle} = \frac{\int_0^{R_n} g_r(P) dr}{\int_0^{R_n} (1 - g_r(P)) dr} \quad (22)$

where $R_n$ is the number of rotations at the chosen value of $g_r(P)$, which should vary from 0.5 to 1.

Therefore,

$$\frac{\langle C_1' \rangle}{\langle C_0' \rangle} = \frac{\left(2\left(\sin^{-1}\left(\sqrt{g_r(P)}\right) - \sin^{-1}\left(\sqrt{P}\right)\right) - \left(\sin\left(2\sin^{-1}\left(\sqrt{g_r(P)}\right)\right) - \sin\left(2\sin^{-1}\left(\sqrt{P}\right)\right)\right)\right)}{\left(2\left(\sin^{-1}\left(\sqrt{g_r(P)}\right) - \sin^{-1}\left(\sqrt{P}\right)\right) + \left(\sin\left(2\sin^{-1}\left(\sqrt{g_r(P)}\right)\right) - \sin\left(2\sin^{-1}\left(\sqrt{P}\right)\right)\right)\right)} \quad (23)$$

Or $\frac{\langle C_1' \rangle}{\langle C_0' \rangle} = \frac{(2X - \sin 2X - (2\theta - \sin 2\theta))}{(2X + \sin 2X - (2\theta + \sin 2\theta))} \quad (24)$

where $X = \sin^{-1}(\sqrt{g_r(P)})$ and $\theta = \sin^{-1}(\sqrt{P})$.

This formula can be employed for setting the ratio of $\langle C_1' \rangle / \langle C_0' \rangle$ for measuring $|S\rangle$ at any probability of finding the target states provided P is known. However, in most real world examples, P is unknown. In order to overcome this limitation, further analysis has been performed by considering the following cases of practical importance to arrive at the rules for assigning values to Set_Val.

Case I: $P = m/N = 1/N$ i.e. $\theta \approx 0$ (for large database)
  (a) $g_r(P) = 0.5$ i.e. $X = \pi/4$
     Inserting value of X and $\theta$ in (24),
     $\langle C_1' \rangle / \langle C_0' \rangle = 0.23$
  (b) $g_r(P) = 0.75$ i.e. $X = \pi/3$
     Inserting value of X and $\theta$ in (24),
     $\langle C_1' \rangle / \langle C_0' \rangle = 0.42$
  (c) $g_r(P) = 1.0$ i.e. $X = \pi/2$
     Inserting value of X and $\theta$ in (24),
     $\langle C_1' \rangle / \langle C_0' \rangle = 1.00$

Case II: $P = m/N = 0.25$ i.e. $\theta = \pi/6$
  (a) $g_r(P) = 0.5$ i.e. $X = \pi/4$
     Inserting value of X and $\theta$ in (24),

Contact: ashish.mani@ieee.org

$\langle C_1' \rangle / \langle C_0' \rangle = 0.60$

(b) $g_r(P) = 0.75$ i.e. $X = \pi/3$

Inserting value of X and θ in (24),

$\langle C_1' \rangle / \langle C_0' \rangle = 1.00$

(c) $g_r(P) = 1.0$ i.e. $X = \pi/2$

Inserting value of X and θ in (24),

$\langle C_1' \rangle / \langle C_0' \rangle = 2.41$

Case III: $P = m/N = 0.5$ i.e. $\theta = \pi/4$

(a) $g_r(P) = 0.5$ i.e. $X = \pi/4$

Inserting value of X and θ in (24),

$\langle C_1' \rangle / \langle C_0' \rangle = 1.00$

(b) $g_r(P) = 0.75$ i.e. $X = \pi/3$

Inserting value of X and θ in (24),

$\langle C_1' \rangle / \langle C_0' \rangle = 1.69$

(c) $g_r(P) = 1.0$ i.e. $X = \pi/2$

Inserting value of X and θ in (24),

$\langle C_1' \rangle / \langle C_0' \rangle = 4.50$

Therefore, in Case I i.e. m = 1, the value of $\langle C_1' \rangle / \langle C_0' \rangle$ should lie between 0.23 and 1.00 for a success probability to lie between 0.5 and 1.0. In Case II i.e. m = N/4, the value of $\langle C_1' \rangle / \langle C_0' \rangle$ should lie between 0.60 and 2.41 for a success probability to lie between 0.5 and 1.0. In Case III i.e. m = N/2, the value of $\langle C_1' \rangle / \langle C_0' \rangle$ should lie between 1.0 and 4.50 for a success probability to lie between 0.5 and 1.0. Thus, if value of **Set_Val is chosen as 1.0** then for all the practical cases (which are represented by Case I to Case III), the probability of success is at least 0.5 for the boundary Case III (for which a classical randomized algorithm would suffice) or greater than 0.5 for all other cases. This is illustrated in Table 1.

TABLE I

SUMMARY OF CALCULATION FOR DETERMINING UNIVERSAL VALUE OF SET_VAL

| Case No. | P = m/N | $g_r(P) = 0.5$ $\langle C_1' \rangle / \langle C_0' \rangle$ | $g_r(P) = 0.75$ $\langle C_1' \rangle / \langle C_0' \rangle$ | $g_r(P) = 1.0$ $\langle C_1' \rangle / \langle C_0' \rangle$ |
|---|---|---|---|---|
| I | 1/N | 0.23 | 0.42 | **1.00** |
| II | 0.25 | 0.60 | **1.00** | 2.41 |
| III | 0.50 | **1.00** | 1.69 | 4.50 |

Contact: ashish.mani@ieee.org

The stopping criterion for the algorithm can be determined directly by measuring the value of $C_1$ and $C_0$ and using (20) to compute $C_1'/C_0'$. The value of $\eta$ is isotropic and universal and a characteristic of the cloning process [11]. Thus, the proposed algorithm is successful and asymptotically as fast as the canonical Grover search algorithm, though, it requires an additional query and local cloning process in each iteration with two extra ancilla qubits.

## IV. CONCLUSION

A fast measurement based fixed-point quantum search algorithm employing disentanglement has been proposed which is better than the existing state of art fixed point algorithms. It is asymptotically as fast as the canonical quantum search algorithm and thus optimal up to a constant factor. It improves canonical quantum search algorithms by making it suitable for search problems with multiple target elements. The proposed technique of using feedback control by measuring ancilla qubits can improve other algorithms that currently use Grover`s search as a subroutine like in optimization etc.

## ACKNOWLEDGMENT

We are extremely grateful to the Most Revered Chairman, Advisory Committee on Education, Dayalbagh, for continued guidance and support in our every endeavor.

Contact: ashish.mani@ieee.org